\documentclass[a4paper,twocolumn,
english,aps,pre,floatfix,showpacs]{revtex4}
\usepackage[T1]{fontenc}
\usepackage[latin1]{inputenc}
\usepackage{amsmath}
\usepackage{babel}
\usepackage{graphics}
\usepackage{amssymb}

\begin{document}
 
\title{Field-driven transition in an Ising magnet with mixed interactions
}

\author{S.L.A. \surname{de Queiroz}}

\email{sldq@if.ufrj.br}

\affiliation{Instituto de F\'\i sica, Universidade Federal do
Rio de Janeiro, Caixa Postal 68528, 21941-972
Rio de Janeiro RJ, Brazil}

\date{\today}
\begin{abstract}
Transfer-matrix methods are used, in conjunction with finite-size scaling and
conformal invariance concepts, to generate an accurate phase diagram for 
a two-dimensional square-lattice Ising spin-1/2 magnet, with couplings
which are positive along one coordinate axis, and negative along the other,
in a uniform external field. Our results indicate that the critical
line starts horizontally at the zero-temperature end of the phase boundary,
at variance with the reentrant behavior predicted in
some earlier studies.
Estimates of the thermal scaling exponent are very close to the Ising value
$y_T=1$ along the critical line, except near $T=0$ where strong crossover
effects prevent a reliable analysis.

\end{abstract}
\pacs{64.60.De, 75.10.Hk, 75.30.Kz}
\maketitle
\section{INTRODUCTION}
\label{intro}

In this paper we investigate a 
square--latice Ising spin--$1/2$ system with ferro-- and 
antiferromagnetic interactions, in the presence of a uniform 
magnetic field. The Hamiltonian is given by:
\begin{equation}
{\cal H}=-J_x \sum_{i,j}\sigma_{i,j}\,\sigma_{i,j+1}
+J_y \sum_{i,j}\sigma_{i,j}\,\sigma_{i+1,j} -H \sum_{i,j} \sigma_{i,j}\ ,
\label{eq:def}
\end{equation}
where $J_x$, $J_y > 0$, and $\sigma_{i,j}=\pm 1$. Here, all fields, 
coupling strengths and temperatures will be given in units of $J_x$. 
At $T=0$, $H < 2 J_y$, the ground state consists of alternating $\pm$ 
stripes  along the $x$ axis, while for larger $H$ all
spins are parallel to the field. A critical line $T_c(H)$ connects
$(T=0,H=2J_y)$ to the zero-field Onsager critical point, which
for $J_y=1$ is at $T_c^{\,0}=2/\ln(1+\sqrt{2})=2.2691853 \dots$.
Since the ground state does not exhibit macroscopic residual entropy, 
the transition along $T_c(H)$ is expected~\cite{cg79} 
to belong to the Ising universality class for all $T \neq 0$,
similarly to the closely-connected case of the standard (isotropic)
antiferromagnet in a uniform field~\cite{mhz77,bdn88,ww90,bw90,wang97b}. 

The problem described by
Eq~(\ref{eq:def}) was treated by M\"uller-Hartmann and Zittartz's 
interface method~\cite{mhz77} in Ref.~\onlinecite{rottman90}, which 
provides an excellent summary of earlier work. It was revisited in
Ref.~\onlinecite{wang97}, using an approach which considers the
zeros of the partition function on an elementary lattice cycle,
and their connection to the free energy singularity at the 
transition~\cite{wang97b}. Remarkably, the critical line found in
Ref.~\onlinecite{wang97} is predicted to
display a positive slope close to  $T=0$, so the critical field 
reaches a maximum at some nonzero $T$ before approaching zero at 
higher temperatures. Similar reentrant behavior was predicted
upon application of a Bethe-Peierls approximation~\cite{prb06}.
On the other hand, the linear-chain approximation~\cite{cg79}
gives an exponentially vanishing positive slope at $T=0$, while the 
interface method also predicts an exponentially vanishing value, only on the
negative side~\cite{rottman90}. Recent real-space renormalization 
results~\cite{dsb09} point to a finite negative slope at $T=0$.

We use transfer-matrix (TM) methods, in connection with finite-size scaling
and conformal invariance ideas, in order to produce a numerically accurate phase 
diagram for this problem. The underlying hypotheses in our work are: (i) that the
phase transition is second-order all along the critical line, and (ii) that
it belongs to the Ising universality class.
Both assumptions are critically reviewed towards the end of the paper, in 
light of the numerical results obtained while assuming their validity.   

In Sec.~\ref{sec:2} we recall the
calculational methods used for the approximate location of the critical line;
the respective results are exhibited, as well as their extrapolation in the 
thermodynamic limit. In Sec.~\ref{conc}, we analyze the data generated in
Sec.~\ref{sec:2}, both in comparison with the existing literature, and
in regard to their internal consistency. The universality of critical behavior 
is discussed, and concluding remarks are made.    

\section{Method and results} \label{sec:2}

We have kept $J_y=1$ in all calculations reported here.
We set up the TM on strips of width $N$ sites, with periodic boundary conditions across.
Referring to Eq.~(\ref{eq:def}), three choices of orientation are available in this 
case, namely the TM can be iterated along the ferromagnetic (F), or $x$-- direction;
along  the  antiferromagnetic (AF), or $y$-- direction; or along the diagonal (D)
of the square lattice. As this is a weakly anisotropic system~\cite{nb83,hucht02},
one would expect estimates of, e.g., critical exponents and locations of
critical points, to converge to the same orientation-independent limit 
for $N \gg 1$, while finite-size corrections should differ in each case.
In order to obey the ground-state symmetry, only even values of 
$N$ are allowed for F and D, while no such restriction applies to AF. 
We generally used $4 \leq N \leq 20$; for F we went up to $N=22$. This range of
$N$ enabled the authors of Ref.~\onlinecite{bw90} to locate the critical line of the
isotropic Ising antiferromagnet to very high accuracy.

\subsection{Keeping $\eta=1/4$}
\label{subsec:2a}

Following earlier work on similar problems~\cite{bdn88,bw90,bww90},
our finite-$N$ estimates for the critical line are found by requiring
that the amplitude-exponent relation of conformal invariance on 
strips~\cite{cardy} be satisfied, with the Ising decay-of-correlations 
exponent $\eta=1/4$: 
\begin{equation}
4 N \kappa_N(T,H)=\pi\ ,
\label{eq:conf-inv}
\end{equation}
where $\kappa_N(T,H)=\ln |\,\lambda_1 /\lambda_2\,|$ 
is the inverse correlation length on a 
strip of width $N$ sites, with $\lambda_1$, $\lambda_2$ being the two largest 
eigenvalues (in absolute value) of the TM.

\begin{figure}
{\centering \resizebox*{3.2in}{!}
{\includegraphics*{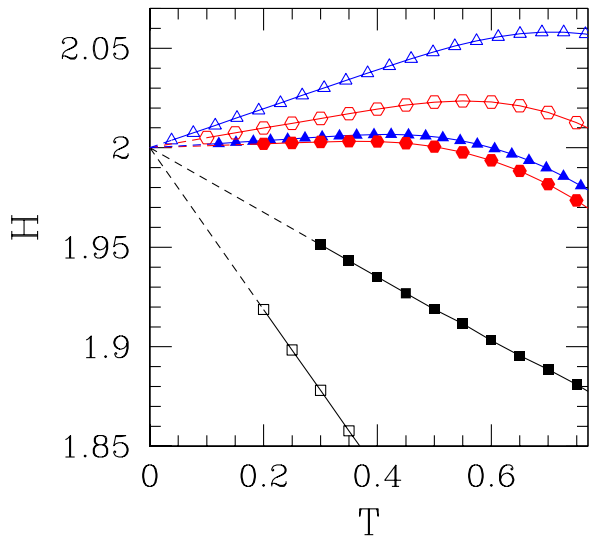}} \par}
\caption{(Color online) Low-temperature approximate critical boundaries given
by solutions of Eq.~({\protect{\ref{eq:conf-inv}}}). Triangles, squares, 
and hexagons denote respectively TM along F, AF, and D direction. Empty symbols:
$N=4$; full symbols: $N=22$ (F), $20$ (D and AF).
}
\label{fig:lowt}
\end{figure}

For $H \ll 1$, the solutions to Eq.~(\ref{eq:conf-inv}) leave the $T$ axis
vertically, and are very close to each other for all three orientations of the TM.
We shall briefly return to this, towards the end of the paper.
For low temperatures $T \lesssim 1$, the main region of interest here,
substantial differences arise. These are illustrated in Fig.~\ref{fig:lowt}. 
One sees that, with increasing $N$, all three families of curves get closer
together, though AF does so at a slower rate. Also, in all cases 
their limiting shape for $T \lesssim 0.4$ is, 
with very good accuracy,
a straight line going through the exact zero-temperature fixed point at $H=2$.     
Reentrant behavior is clearly visible for the F curves and, to a lesser
extent, for the D ones, though in both cases the peaks turn flatter as $N$ 
increases. In order to check whether the reentrancies vanish in the $N \to 
\infty$ limit, we plotted the sequences of slopes $S_N$ of the $T \to 0$
straight-line sections referred to above, against $N^{-1}$. 

\begin{figure}
{\centering \resizebox*{3.2in}{!}
{\includegraphics*{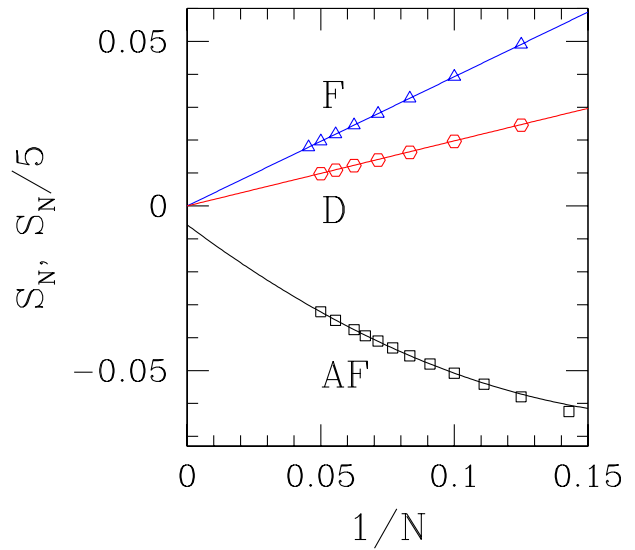}} \par}
\caption{(Color online) Slopes $S_N$ of the low-temperature straight-line
sections of approximate critical boundaries 
given by solutions of Eq.~({\protect{\ref{eq:conf-inv}}}), against $1/N$. 
Triangles, squares, and hexagons denote respectively TM along F, AF, and D direction. 
For AF, data are scaled by a factor of 5. Lines are fits to data: linear for F and D
(all $N$), and parabolic
(linear plus quadratic)
for AF ($N \geq 10$ only). 
}
\label{fig:slope}
\end{figure}

Results are shown in 
Fig.~\ref{fig:slope}. For both F and D the trend is, to an excellent approximation,
$S_N \to 0$ linearly with $N^{-1}$. Indeed, the respective extrapolated 
values are $(4 \pm 3)\times 10^{-7}$ (F) and $(-15 \pm 4)\times 10^{-5}$ (D),
which in practice equate to zero within the present context.
For AF, a large amount of curvature is present; from a parabolic fit
($S_N=S_\infty+a\,N^{-1}+b\,N^{-2}$)
of $N \geq 10$ data, one gets $S_\infty =-0.029 \pm 0.004$.
This would correspond to $1-H_c(T)/H_c(0) \simeq 0.01$ 
at $T=0.5$ which, though small, is significant. 

In summary, the analysis of limiting slopes at low $T$,  $N \to \infty$,
indicates that the reentrant behavior observed for F and D data
is a finite-size effect, which tends to
vanish in the thermodynamic limit. A small discrepancy still remains, 
between F and D data which are consistent with a horizontal critical 
line at $T \to 0$, and AF results which point to a slightly negative slope
in that limit.

\subsection{Phenomenological Renormalization}
\label{subsec:2b}

Further insight can be gained by relaxing the assumption made in 
Eq~(\ref{eq:conf-inv}), that the phase transition belongs to the Ising
universality class, and demanding only that it remain of second order.  
From finite-size scaling, one gets the basic equation of the
phenomenological renormalization group (PRG)~\cite{fs2} for the critical line: 
\begin{equation}
N \kappa_N(T,H)=N^{\prime} \kappa_{N^{\prime}}(T,H) ,
\label{eq:prg}
\end{equation}
where the strip widths $N$ and $N^{\prime}$ are to be taken as close as
possible for improved convergence of results against increasing $N$. This means 
$N^{\prime}=N-2$ for F, D, and $N-1$ for AF.

We found that PRG results converge very rapidly for both F and D orientations 
of the TM,  without any sign of the reentrances shown by the solutions
of Eq.~(\ref{eq:conf-inv}). For the F case, the largest discrepancies  
between $N=6$ and $8$ amount to $0.8\%$ close to $H=0$, and are slowly
reduced upon increasing $H$; around $T \simeq 0.7$, $H \simeq 1.97$,
the two curves differ by $0.04\%$. At $T=0.4$, both
coincide to within two parts in $10^5$, at less than $0.1\%$ of $H=2$, 
and then home in towards $(T,H)=(0,2)$ on a straight line.
The discrepancy between $N=12$ and $14$ is never more than one
part in $10^4$ for $T \leq 1$.
The picture is quantitatively similar for the D orientation.    

On the other hand, for PRG with the TM along the AF direction, one gets
relatively large negative slopes (but approaching zero with increasing $N$)
as $T \to 0$. Furthermore, at intermediate temperatures $0.4 \lesssim T
\lesssim 0.8$ the curves show inflection points which make extrapolation
of such sections to $N \to \infty$ prone to instabilities.

\begin{figure}
{\centering \resizebox*{3.2in}{!}
{\includegraphics*{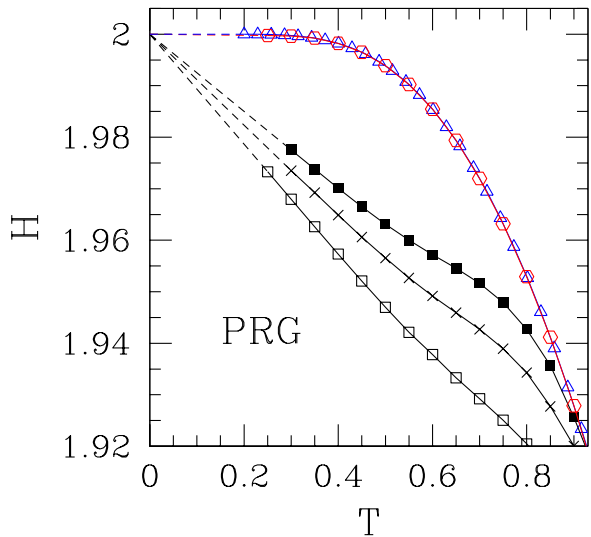}} \par}
\caption{(Color online) Low-temperature approximate critical boundaries given
by solutions of Eq.~({\protect{\ref{eq:prg}}}). Triangles: TM along F direction,
$N=14$;  hexagons: D, $N=14$; empty squares, crosses, and full squares all for AF,
respectively $N=10$, $12$, and $14$.  
}
\label{fig:prg}
\end{figure}
\begin{figure}
{\centering \resizebox*{3.2in}{!}
{\includegraphics*{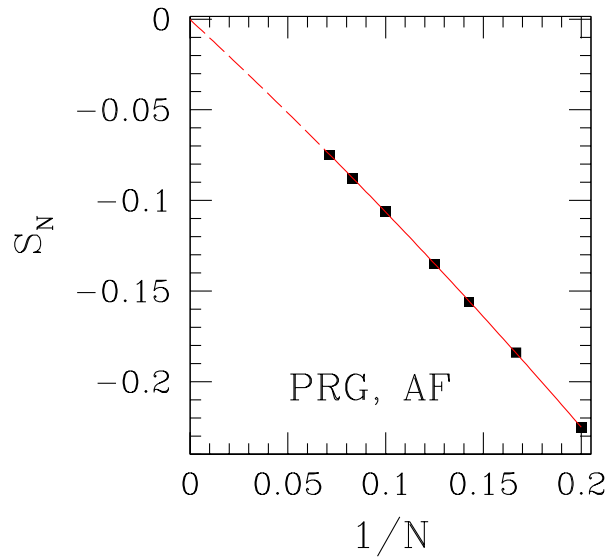}} \par}
\caption{(Color online) Slopes $S_N$ of the low-temperature straight-line
sections of approximate critical boundaries 
given by the solution of Eq.~({\protect{\ref{eq:prg}}}), for TM in the AF 
direction, against $1/N$. Data points are for $N=5-14$.
The line is a parabolic 
(linear plus quadratic)
fit to data.
}
\label{fig:slope_afprg}
\end{figure}
The above-mentioned features are illustrated in Fig.~\ref{fig:prg}. We
investigated the behavior of the limiting slopes for the AF family against
$1/N$, with results displayed in Fig.~\ref{fig:slope_afprg}. 
The line shown in the Figure is a parabolic fit to data
($S_N=S_\infty+a\,N^{-1}+b\,N^{-2}$),
from which one gets $ |S_\infty| <10^{-3}$, i.e., essentially zero.

\subsection{Extrapolations}
\label{subsec:2c}

In this Subsection, we deal directly with extrapolations of finite-size
data for all $(T,H)$. This is in contrast to the analysis of slopes, which
applies only at low $T$ where the finite-$N$ data actually fall on straight  
lines.

Extensive investigation of the related problem of isotropic Ising antiferromagnets 
in a field~\cite{bdn88} shows that, in that case, the main irrelevant
exponent is $y_{\rm ir}=-2$, i.e., the leading corrections to scaling are expected to
depend on $N^{-2}$. Thus, it is plausible to assume that
such corrections also play a dominant role here. However, the 
analysis of slopes above suggests that corrections in $N^{-1}$ are
present as well, at least in the low-temperature region. 

In what follows, we examine both scenarios, i.e., $y_{\rm ir}=-1$ and $-2$.
The locations of points on the approximate critical line, for strip width $N$, 
are denoted by $(T_N^\ast,H_N^\ast)$.
On account of the overall shape of the phase diagram, we
extrapolate against negative powers of $N$ in two different ways: (i) 
at constant $X=H$ for high $T \gtrsim 1.7$, and (ii) 
at constant $X=T$ for $T \lesssim 1.7$. For large $N$, we
fit $Y_N$ [$=T_N^\ast(H)$ in (i), or $H_N^\ast(T)$ in (ii)] to a form
\begin{equation}
Y_N^\ast(X)=Y_{\rm ext}(X) +a_{\,y_{\rm 
ir}}(X)\,N^{\,y_{\rm ir}}+b_{\,y_{\rm ir}}(X)\,N^{2y_{\rm ir}}
\ .
\label{eq:fs1}
\end{equation}  
We have chosen to calculate $Y_{\rm ext}(X)$, $a_{\,y_{\rm ir}}(X)$, and 
$b_{\,y_{\rm ir}}(X)$
from the three largest values of $N$ available for each strip orientation, 
thus error bars (other than those associated to the positions $Y_N^\ast(X)$
themselves) are not available from this procedure. 

We managed to produce well-behaved extrapolated critical lines from 
the solutions to Eq.~(\ref{eq:conf-inv}), only for F and D orientations
of the TM. For AF, though extrapolations are generally smooth for 
$T \gtrsim 0.6$,
they display instabilities for lower $T$, i.e. the section of the phase diagram
which is most relevant in the search for reentrant behavior.
As regards PRG curves, due to the  fast convergence of F and D data
we found that $N=14$ data can already be taken as very close to the $N \to \infty$
limit, to within an estimated one part in $10^4$. As mentioned above,
the extrapolated PRG curves for AF display instabilities, except for
very low $T$ (for which the relevant information is summarized in the
slope analysis illustrated in Fig.~\ref{fig:slope_afprg}).

\begin{figure}
{\centering \resizebox*{3.2in}{!}
{\includegraphics*{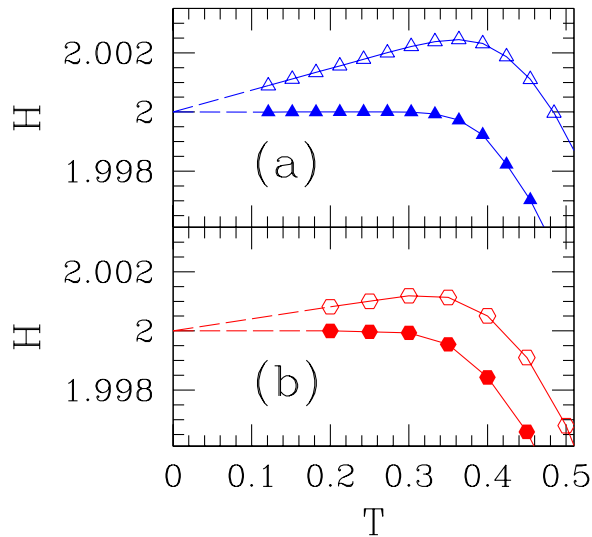}} \par}
\caption{(Color online) Low-temperature extrapolated critical boundaries, 
obtained via Eq.~({\protect{\ref{eq:fs1}}}), with $y_{\rm ir}=-2$ (empty symbols),
and $y_{\rm ir}=-1$ (full symbols). TM is in the F direction in (a), and along
the diagonal (D) in (b).
}
\label{fig:extr}
\end{figure}

Fig.~\ref{fig:extr} shows the low-temperature regions of extrapolated critical
boundaries, for F and D, assuming $y_{\rm ir}=-1$ or $-2$ in Eq.~(\ref{eq:fs1}).
For both curves corresponding to $y_{\rm ir}=-1$, there is a rather flat section:
for F, points with $T \leq 0.36$ remain within one part in $10^5$ from $H=2$,
while for D the $T \leq 0.35$ region is within one part in $10^4$ from that
limit. On the other hand, the $y_{\rm ir}=-2$ extrapolations exhibit tiny reentrances,
with maxima respectively at $H=2.0024$ (F) and $H=2.0012$ (D).

In order to get further insight into the competing scenarios under 
investigation, we examine the behavior of the coefficients $a_{-1}$ and 
$b_{-1}$ of Eq.~(\ref{eq:fs1}) along the critical line. Indeed, this 
amounts to an unbiased test of whether the dominant
$N$- dependence of finite-size data is on $N^{-1}$ (via $a_{-1}$) 
or $N^{-2}$ (via $b_{-1}$). 
\begin{figure}
{\centering \resizebox*{3.2in}{!}
{\includegraphics*{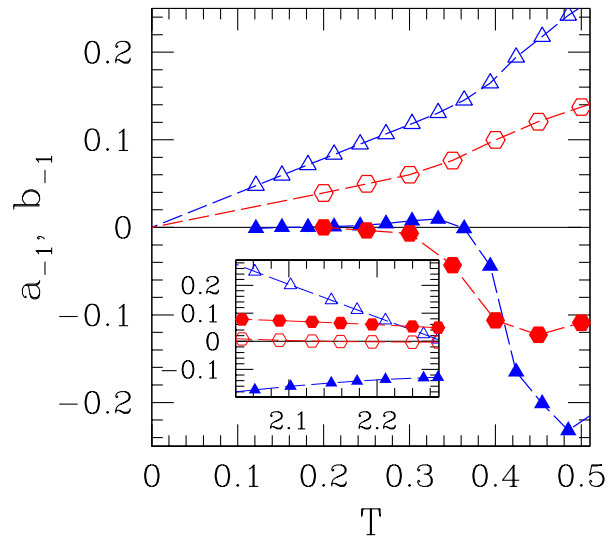}} \par}
\caption{(Color online) Low-temperature behavior of fitting coefficients 
$a_{-1}$ (empty symbols), $b_{-1}$ 
(full symbols) of Eq.~({\protect{\ref{eq:fs1}}}). Triangles: TM along F direction.
Hexagons: TM along diagonal (D). Inset: high-temperature behavior. 
Same axis labels and symbol captions as in main figure (see text).
}
\label{fig:coeffs}
\end{figure}
In Figure~\ref{fig:coeffs} one sees that, at low $T$, the $N^{-2}$ terms
tend to vanish for both F and D cases. Furthermore, the inset of the 
Figure shows that in the low-field limit, it is the $N^{-2}$ corrections
that become dominant, and those in  $N^{-1}$ become negligible as $H \to 0$, 
as is well known for the zero-field Ising model~\cite{bdn88,dQ00}. 
The latter result gives further credence to the procedure just described, 
and thus to the conclusions reached regarding the low--$T$ regime.

Finally, we found no clear evidence that an exponent $y_{\rm vac}=-4/3$,
associated to vacancy excitations, might be present~\cite{bdn88}. This is 
because
(i) at the F and D data clearly depend predominantly on $N^{-1}$ at low $T$,
as already shown in the slope analysis; and (ii) the slope data for AF do not 
exhibit any significant improvement in quality of fit when plotted against
$N^{-4/3}$ (compared to the parabolic fits in $N^{-1}$ of Figs.~\ref{fig:slope}
and~\ref{fig:slope_afprg}).

\section{Discussion and conclusions} \label{conc}

We begin by recalling that Eq.~(\ref{eq:prg}) depends on the existence of an 
underlying diverging correlation length, thus its non-trivial fixed points 
correspond to a second-order transition. We only failed to find such 
points for low temperatures, generally $T \lesssim 0.2$. This is due to 
the extremely slow convergence of numerical results,  connected to the 
very large ratio between positive and negative exponentials which are 
the TM states' Boltzmann weights in that region. If a tricritical point
were present, separating first- and second-order sections of the critical 
curve, one would expect spurious effects such as the "hooking" found for 
the three-dimensional version of the current problem~\cite{hs72}. 
On the contrary, as long as we can find low--$T$ solutions to  
Eq.~(\ref{eq:prg}) they behave in the expected manner, i.e.  
homing in towards the exact $T=0$ fixed point at $H=2$.
Furthermore, the extrapolated $\eta=1/4$ curves agree very well 
with the solutions of Eq.~(\ref{eq:prg}) down to $T=0.2$, and 
still extend somewhat further down to $T \approx 0.1$
Since, by conformal invariance, $\eta=1/4$ corresponds
to an Ising transition, our results indicate that this is the character
of the critical line, at least down to $T=0.1$. Therefore, if a 
tricritical  point is present, it must be located at $T < 0.1$, $H \approx 
2$. We thus conclude that the 
transition is indeed second-order and in the Ising universality class
along the whole of the critical curve (except for the latter possibility,
which we are not able to probe directly). 

We now recall that the reentrant behavior predicted in 
Ref.~\onlinecite{wang97}
is sizable: in the units used in the present work, it translates into
the critical line leaving  $(T,H)=(0,2)$ with a slope $S=(1/2)\ln 2 =0.3466 \dots$
[$\,$see their Eq.~(31)$\,$]. This is in contrast with the results of
Subsec.~\ref{subsec:2c}, where we find at most $S=7 \times 10^{-3}$
(with the TM in the F direction, using $y_{\rm ir}=-2$).

In the comparable problem of isotropic antiferromagnets, though the 
critical line $H_c(T)$ given in Ref.~\onlinecite{wang97b} does not 
exhibit reentrances, it is always above that found in 
Refs.~\onlinecite{ww90,bw90} (except at the $T=0$ and $H=0$ ends,
where both lines coincide). The maximum discrepancy, of order $4\%$,
is in the central  region, $0.7 \lesssim T \lesssim 1.5$, tailing off
towards both ends. It thus appears that the methods employed in
Refs.~\onlinecite{wang97b,wang97} generally tend to overestimate the 
extent of the ordered region in parameter space.

Though the present problem is weakly anisotropic in a broad sense,
the  distinct nature of spin couplings along each coordinate 
axis is responsible for the introduction of subtle biases, when one 
iterates the TM along either of those very same axes.
Indeed, the size of the ordered region predicted 
by low-$T$  TM results systematically
decreases as one changes orientation from F to D, and finally to AF. 
The explanation is that, for the high fields near the 
critical curve, the ferromagnetic correlations picked out when the TM 
goes along F tend to be emphasized; when the TM goes along AF,
the corresponding antiferromagnetic correlations are inhibited
by the field. The small negative extrapolated 
slope of the AF curves, shown in Fig.~\ref{fig:slope}, most 
likely reflects the latter effect.
For the TM going along D, an evenly-balanced
mixture of both kinds of correlations is collected upon its iteration.
Therefore we believe that, of the three setups implemented
here, the latter is the one likely to produce the most reliable
results. 

Note also that, for PRG one is always comparing correlation lengths 
evaluated along the same lattice direction in Eq.~(\ref{eq:prg}), 
so the aforementioned biases tend to cancel out, if present. 
Thus, (i) the PRG curves for F  and D do not show reentrancies, 
even for finite $N$; and (ii) the slope
of PRG curves for AF approaches zero as $N^{-1} \to 0$ (see 
fig.~\ref{fig:slope_afprg}), as opposed to the small negative
value found for the corresponding solutions of Eq.~(\ref{eq:conf-inv}).

In principle, the results for F, D, and AF orientations must eventually 
extrapolate to the same location of the critical
boundary as $N \to \infty$. The way finite--$N$ data vary as $N$ increases 
is consistent with this, see Figs.~\ref{fig:lowt} and~\ref{fig:prg}.
However, it is apparent that subdominant corrections to scaling
have much larger amplitude for AF than for F or D. Attempting a
theoretical understanding of why this is so usually becomes a highly 
nontrivial task, as it involves (i) unequivocally identifying the 
associated irrelevant exponents, and (ii) once this is done, 
analysing the (non-universal) amplitudes of the corresponding terms.
Examples of this can be seen in Refs.~\onlinecite{bdn88} 
and~\onlinecite{dQ00}. Here, for practical reasons we chose to 
extrapolate only the sets of $(T,H)$ data for which the 
small subdominant corrections could be satisfactorily dealt
 with via Eq.~(\ref{eq:fs1}) and its underlying assumptions.  
 
\begin{figure}
{\centering \resizebox*{3.2in}{!}
{\includegraphics*{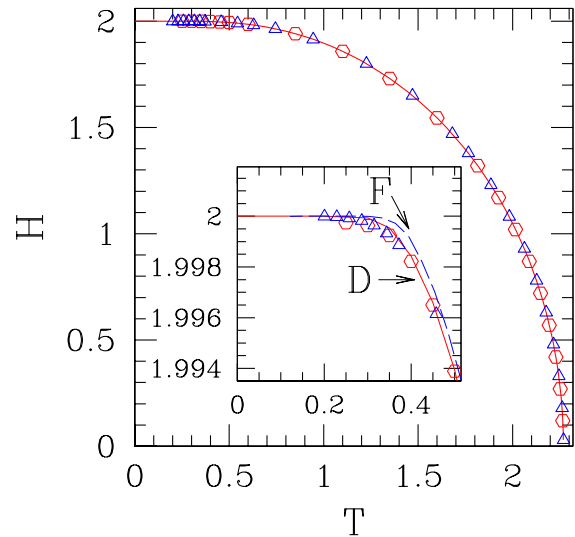}} \par}
\caption{(Color online) General view of extrapolated phase diagram.
Full line is extrapolation of solutions of 
Eq.~({\protect{\ref{eq:conf-inv}}}),
with the TM along D, using $y_{\rm ir}=-1$. Points: PRG, $N=14$,
with the TM along F (triangles) and D (hexagons).
Inset: Low-temperature section of same data. 
Same axis labels and symbol captions as in main figure,
except for additional dashed line which is
extrapolation of solutions of Eq.~({\protect{\ref{eq:conf-inv}}}),
with the TM along F, using $y_{\rm ir}=-1$.
}
\label{fig:extr2}
\end{figure}

We have found that the minimum amount of discrepancy among all
our results corresponds to the set of: (i) extrapolated data
from the solutions of Eq.~(\ref{eq:conf-inv}), with the TM along D, 
using $y_{\rm ir}=-1$; (ii) PRG with the TM along D, and (iii)
PRG with the TM along F. See Fig.~\ref{fig:extr2}.
The extrapolations from the solutions of 
Eq.~(\ref{eq:conf-inv}), with the TM along F, and using $y_{\rm ir}=-1$,
also agree very well with these, except for  the low-temperature 
'shoulder' where curves begin to depart more significantly from
the horizontal line $H=2$. This is illustrated in the inset of
Fig.~\ref{fig:extr2}.
\begin{figure}
{\centering \resizebox*{3.2in}{!}
{\includegraphics*{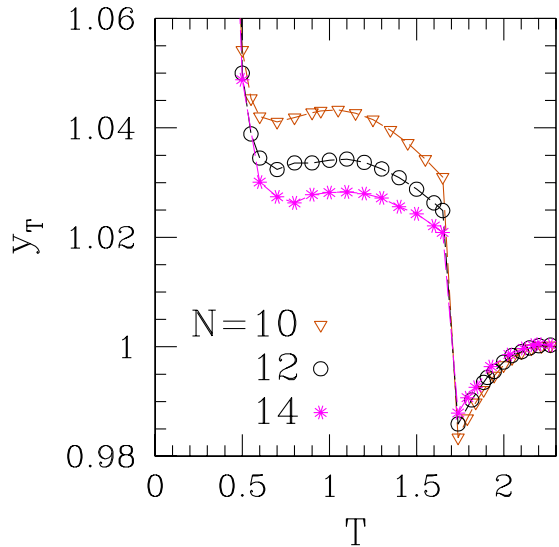}} \par}
\caption{(Color online) Thermal scaling exponent $y_T=1/\nu$ along extrapolated 
critical line shown in Fig.{\protect{\ref{fig:extr2}}}, calculated via 
Eq.~({\protect{\ref{eq:y_prg}}}), with $M=N-2$. TM along D. The discontinuity
at $T \approx 1.74$ marks the change in assumed scaling direction (see text). 
}
\label{fig:yt}
\end{figure}

In order to check on the universality of critical properties, we examined the
thermal exponent $y_T=1/\nu$ along the critical line. Considering two strips
of widths $M$ and $N$, finite-size scaling~\cite{fs2} gives:
\begin{equation}
y_T=1+\frac{\ln (\kappa_N^\prime/\kappa_M^\prime)}{\ln (M/N)}\ ,
\label{eq:y_prg}
\end{equation} 
where $\kappa_N^\prime$, $\kappa_M^\prime$ are derivatives of the inverse 
correlation lengths, taken with respect to
the appropriate temperature-like scaling field, evaluated on the critical curve.
With the TM along D, and using $M=N-2$, we swept the extrapolated line shown in
Fig.~\ref{fig:extr2}. For simplicity, the temperature-like direction was taken
as the temperature axis for low and middle fields $H <1.41$ 
(corresponding to $1.74 \lesssim T \leq 2.269\dots $), and as the $H$-axis for the 
remainder
of the critical line. The resulting solutions to 
Eq.~(\ref{eq:y_prg}), for $N=10$, $12$, and $14$, 
are shown in Fig.~\ref{fig:yt}. Although the abrupt jump at $T \approx 1.74$
is an artifact, reflecting the above-mentioned (and somewhat arbitrary) change 
in the assumed scaling direction, one sees that on both sides of the
discontinuity the estimates are rather close to the Ising value $y_T=1$,
and systematically approach it with increasing $N$. On the other hand,
for $T \lesssim 0.5$ (where $H$ already differs by less than $0.5\%$
from the zero-temperature $H_c(0)=2$), crossover effects related to the
energy level crossings at $T=0$ cause an extreme deterioration in our 
estimates.   
 
Near the $H=0$ extreme of the critical curve, we have
fitted our extrapolated curves to a parabolic shape:
\begin{equation}
T_c(H)=T_c(0)-a\,H^{2}\ ,
\label{eq:tc0}
\end{equation}
from which we get $a=0.217\pm 0.001$, to be compared with 
$a=0.1767\dots$ and $a=0.3018\dots$, each coming from a slightly 
different implementation of the interface method~\cite{rottman90}.
Note that the coefficient of a hypothetical linear term in 
Eq.~(\ref{eq:tc0}) vanishes identically because the phase diagram is 
symmetric under field inversion, thus the scaling variable must 
depend on $H^2$. For $H \to 0$ this means that $T_c(H)=T_c(0)-aH^2$
plus higher-order terms~\cite{mk87}. 

Finally, we return to the exponentially vanishing deviations from a
horizontal line near $T=0$, predicted in earlier work and
mentioned in the Introduction.
These are of the general form~\cite{rottman90}
\begin{equation}
H=2+c\,T^x\,\exp(-d/T)\ ,
\label{eq:expdev}
\end{equation}
where $c<0$ for both implementations of the interface 
method~\cite{rottman90}, as well as for a free-fermion 
approximation~\cite{rottman90}, whereas $c>0$ for the
linear-chain approximation~\cite{cg79,rottman90} ($|c|$, $d$, and $x$
turn out to be of order unity in all cases). 
While we do not have enough accuracy at low temperatures to probe for this
sort of effect, it seems safe to state that any stronger deviations
from the horizontal, be they in the shape of a reentrance or the 
opposite, are ruled out by our results.

Our extrapolated data for the location of the critical line,
for both F and D, with $y_{\rm ir}=-1$ and $-2$,  
are available as ASCII files~\cite{epaps}.
%

\begin{acknowledgments}
This research was partially supported by the Brazilian agencies CNPq
under Grant~No.~30.6302/2006-3, FAPERJ under Grants Nos. 
E26--100.604/2007 and E26--110.300/2007, and CAPES.
\end{acknowledgments}

\end{document}